\documentclass[times, review, 10pt]{elsarticle}

\usepackage{amssymb}
\usepackage{amsmath}

\usepackage{multirow,subfigure, amsfonts,bbding, url}

\journal{Pattern Recognition}

\begin{document}

\begin{frontmatter}



\title{
Audio-Visual Representation Learning via Knowledge Distillation from Speech Foundation Models}


\author[snnu]{Jing-Xuan Zhang\corref{cor1}} 
\ead{jxzhanggg@snnu.edu.cn}

\author[ustc,ifly]{Genshun Wan} 
\ead{gswan@iflytek.com}

\author[ifly]{Jianqing Gao} 
\ead{jqgao@iflytek.com}

\author[ustc]{Zhen-Hua Ling} 
\ead{zhling@ustc.edu.cn}

\cortext[cor1]{Corresponding author}

\affiliation[snnu]{organization={School of Computer Science, Shaanxi Normal University},
            addressline={No. 620, West Chang'an Avenue}, 
            city={Xi'an},
            postcode={710119}, 
            state={Shaanxi},
            country={P.R.China}}

\affiliation[ustc]{organization={University of Science and Technology of China},
            addressline={No.96, JinZhai Road}, 
            city={Hefei},
            postcode={230026}, 
            state={Anhui},
            country={P.R.China}}

\affiliation[ifly]{organization={iFLYTEK Research, iFLYTEK Co. Ltd.},
            addressline={No. 666, Wangjiang West Road}, 
            city={Hefei},
            postcode={230088}, 
            state={Anhui},
            country={P.R.China}}


\begin{abstract}

Audio-visual representation learning is crucial for advancing multimodal speech processing tasks, such as lipreading and audio-visual speech recognition.
Recently, speech foundation models (SFMs) have shown remarkable generalization capabilities across various speech-related tasks.
Building on this progress, we propose an audio-visual representation learning model that leverages cross-modal knowledge distillation from SFMs.
In our method, SFMs serve as teachers, from which multi-layer hidden representations are extracted using clean audio inputs.
We also introduce a multi-teacher ensemble method to distill the student, which receives audio-visual data as inputs.
 A novel representational knowledge distillation 
loss is employed to train the student during pretraining,
which is also applied 
during finetuning to further enhance the performance 
on downstream tasks.
Our experiments utilized both a self-supervised SFM, WavLM, and a supervised SFM, iFLYTEK-speech.
The results  demonstrated 
that our proposed method achieved  superior  or at least comparable performance to previous state-of-the-art baselines across automatic speech recognition, visual speech recognition, and audio-visual speech recognition tasks. Additionally, comprehensive ablation studies and the visualization of learned representations were conducted to evaluate the effectiveness of our proposed method.

\end{abstract}




\end{frontmatter}


\section{Introduction}
\label{sec:intro}


Speech processing techniques have advanced significantly in recent years, with speech foundation models (SFMs) emerging as a prominent development. Trained on vast speech datasets, these large-scale models such as WavLM~\cite{chen2022wavlm} and Whisper~\cite{radford2023robust}, excel in numerous tasks. However, environmental noise remains a critical challenge, potentially compromising the performance of audio-based models. 
This limitation has spurred interest in audio-visual speech processing. Visual speech recognition (VSR), as known as lipreading~\cite{SHIN2011559}, focuses on decoding speech content from visual cues alone, while audio-visual speech recognition (AVSR)~\cite{afouras2018deep} integrates both auditory and visual inputs. 
These multi-modal approaches contribute to more robust speech recognition systems, showing promise for enhancing communication in noisy environments and improving hearing aids.


Audio-visual representation learning underpins contemporary advancements in audio-visual speech processing. 
In contrast to early work using handcrafted features, deep neural networks now extract useful representations by leveraging large-scale datasets.
The training objectives 
can be broadly  categorized into supervised learning and self-supervised learning. While supervised learning requires labeled data, self-supervised learning utilizes unlabeled data to extract general representations during pretraining phase~\cite{YU2024110016}. Subsequently, the model can be adapted to a wide-range of downstream tasks with minimal labeled data. 
Various pretext tasks have been introduced for audio-visual self-supervised learning (AV-SSL), including mask prediction~\cite{shiavhubert,zhu2023vatlm}, contrastive learning~\cite{NEURIPS2021_38ef4b66,zhang2022learning}, and cross-modal distillation~\cite{haliassos2022jointly,Lian2023avdata2vec}. 
Despite their success, AV-SSL models face several limitations.
First, collecting parallel audio and corresponding lip video data is more challenging than acquiring audio data alone. Consequently, current AV-SSL models are optimized with audio-visual datasets that are substantially smaller than those used to train SFMs.
Second, AV-SSL methods often rely on iterative training~\cite{shiavhubert,hsu2022u} or an online teacher module~\cite{haliassos2022jointly,Zhang2023av2vec} to generate the learning targets
 for pretext tasks, which can introduce additional computational overhead or potentially lead to unstable training.

To address the aforementioned limitations and inspired by the 
success of speech foundation models,
we propose transferring the knowledge from SFMs to a joint audio-visual representation learning model through cross-modal knowledge distillation.
In our method, the SFM serves as the teacher,
with its multi-layer hidden representations extracted and aggregated to distill knowledge into the student model.
 An ensemble of multiple teachers can further  enhance the student's generalization
capability.
The student is exposed to noisy audio-visual data to improve its robustness to noise.
We introduce a novel representational knowledge distillation loss that combines both feature regression and KL divergence loss using soft cluster labels. Furthermore, the pretrained model is finetuned with an auxiliary knowledge distillation loss to enhance performance on downstream tasks, including Automatic Speech Recognition (ASR), VSR, and AVSR.
Our work is closely related to LiRA~\cite{ma2021lira}, which learns visual speech representations through self-supervision of audio representation extracted by PASE+~\cite{ravanelli2020multi}. 
The key distinction between our approach and LiRA lies in our focus on exploiting knowledge learned from large-scale audio data. Consequently, we utilize SFMs trained on substantially larger audio datasets compared to PASE+. Moreover, our method jointly learns audio-visual representations within a single model.

In our experiments, we  evaluated two speech foundation models: a self-supervised model, WavLM~\cite{chen2022wavlm}, and a supervised model, iFLYTEK-speech.
Our method demonstrated superior or at least comparable performance to baselines across ASR, VSR, and AVSR tasks, particularly in scenarios with low-resource labeled data. 
Additionally, we compared our method to baseline models using noisy audio-visual test sets in the AVSR task.
Comprehensive ablation studies were conducted to validate the effectiveness of our approach, focusing on key components such as
multi-teacher ensemble, our proposed knowledge distillation loss, and the joint modeling of audio-visual speech in a single student model. 
Moreover, we performed a visualization analysis of the representations derived from both audio and visual inputs. Our findings revealed a notable alignment between the audio and visual representations, providing further support for the recently observed trend of representational convergence across modalities~\cite{hsu2022u}.



\section{Related works}
\subsection{Knowledge distillation}
Knowledge distillation (KD) has emerged as a crucial technique in model compression and transfer learning.
Cross-modal knowledge distillation (CMKD), an extension of KD, focuses on transferring knowledge between different modalities~\cite{10643687}.
Many studies have explored cross-modal distillation for audio-visual speech data, leveraging the inherent synergy and correspondence between audio signals and lip movements. 
Afouras et al.~\cite{afouras2020asr} train a lipreading model by distillation from an ASR model, combining connectionist temporal classification with frame-wise cross entropy loss. 
LIBS~\cite{zhao2020hearing} enhances lipreading by distilling multi-scale knowledge from speech recognizers. 
Another line of research employs CMKD within a self-supervised learning framework. 
RAVEn~\cite{haliassos2022jointly} utilizes an asymmetric distillation scheme for audio-visual representation learning with modality-specific student. 
Both AV-data2vec~\cite{Lian2023avdata2vec} and our previously proposed AV2vec~\cite{Zhang2023av2vec} adopt a multimodal teacher for distilling a student that accepts audio, visual, and audio-visual inputs.


Our work is also related to research exploring ensemble-based approaches in knowledge distillation. 
For instance, Huang et al.~\cite{10096445} propose an ensemble knowledge distillation framework to transfer knowledge from large self-supervised speech models. 
Similarly, recent studies~\cite{10605999,10506099} develop a dynamic ensemble teacher-student architecture for fake audio detection. 
Building upon these works, our approach leverages aggregated representations from multiple teacher networks across different layers to enhance knowledge distillation.


\subsection{Visual and audio-visual speech recognition}

Nowadays, end-to-end frameworks have become the predominant approach for
visual and audio-visual speech recognition~\cite{afouras2018deep,ma2021end}, where a single network consumes raw lip video pixels and audio features to generate text outputs. The end-to-end paradigm relies on sequence-to-sequence model architectures and their corresponding loss function, including encoder-decoder with attention (ED)~\cite{afouras2018deep}, connectionist temporal classification (CTC)~\cite{petridis2018audio}, and neural transducer~\cite{makino2019recurrent}. 
Under this framework, researchers elaborate on designing neural network architecture for extracting visual representations, such as
 temporal convolutional network (TCN)~\cite{martinez2020lipreading}, Transformer~\cite{SerdyukBS22}, visual Transformer pooling~\cite{prajwal2022sub}, and Conformer~\cite{ma2021end}.
Concurrently, another line of research explores effective audio-visual fusion strategy for AVSR, such as V-CAFE~\cite{hong2022visual}, and unified cross-modal attention~\cite{li2024unified}. 


Recently, there has been a growing trend in leveraging large-scale ASR models for visual and audio-visual speech recognition. 
Djilali et al.~\cite{djilali2023lip2vec} pioneered the reprogramming of ASR models for VSR by introducing a prior network that transforms visual representations into audio representations. 
Building on a similar concept, Prajwal et al.~\cite{prajwal24_interspeech} proposed a VSR approach with a simplified training objective. 
In the AVSR domain, Simic et al.~\cite{10448047} enhanced the Whisper model, originally designed for ASR, by incorporating a fusion module for audio and visual inputs. 
Similarly, Whisper-flamingo~\cite{rouditchenko24_interspeech} extended the Whisper model's capability through additional cross-attention components that process visual representations for AVSR tasks. 
More recently, the integration of large language models with pretrained audio and visual encoders has achieved state-of-the-art performance in AVSR tasks~\cite{cappellazzo2024large}. 
Our work distinguishes itself by learning general audio-visual representations applicable beyond conventional lipreading and AVSR tasks. 
Moreover, our approach can be seamlessly integrated with these existing techniques to further enhance 
visual and audio-visual speech recognition performance in future work.


\subsection{Speech foundation models}

Foundation models, a term first coined by Bommasani et al.~\cite{Bommasani2021FoundationModels}, are defined as large models trained on broad data that adapts to a wide range of downstream tasks. Various speech foundation models have emerged, broadly categorized into two types, i.e. supervised learning-based and self-supervised learning (SSL)-based models.
Among SSL-based speech foundation models, 
HuBERT~\cite{hsu2021hubert}  utilizes k-means clustering to generate pseudo-class labels for model training using a mask prediction loss,
with iterative refinement through subsequent clustering and mask prediction steps.
WavLM~\cite{hsu2021hubert} improves HuBERT by using a more diverse pretraining dataset and performs speech denoising modeling during pretraining.
On the other hand, Whisper family~\cite{radford2023robust} exemplifies  supervised learning-based speech foundation model. Trained on large-scale multilingual and multitask labeled data, it demonstrates robust performance in ASR and speech translation tasks. 
In our experiments, we evaluated both a self-supervised speech foundation model, WavLM, and a supervised one, iFLYTEK-speech.

\begin{figure}
    \centering
    \includegraphics[width=0.9\linewidth]{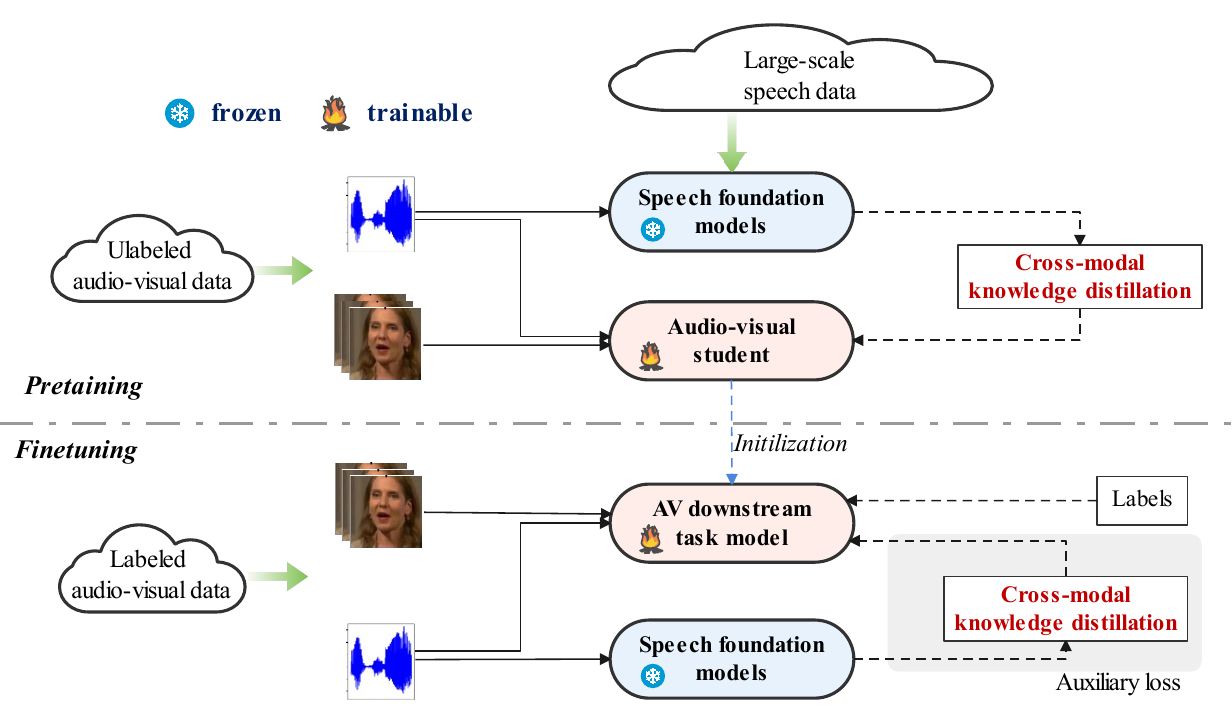}
    \caption{Illustration of  overall scheme of our proposed method during the pretraining and finetuning phases.}
    \label{fig:figure1}
\end{figure}

\section{Proposed method}

\begin{figure}
    \centering
    \includegraphics[width=0.9\linewidth]{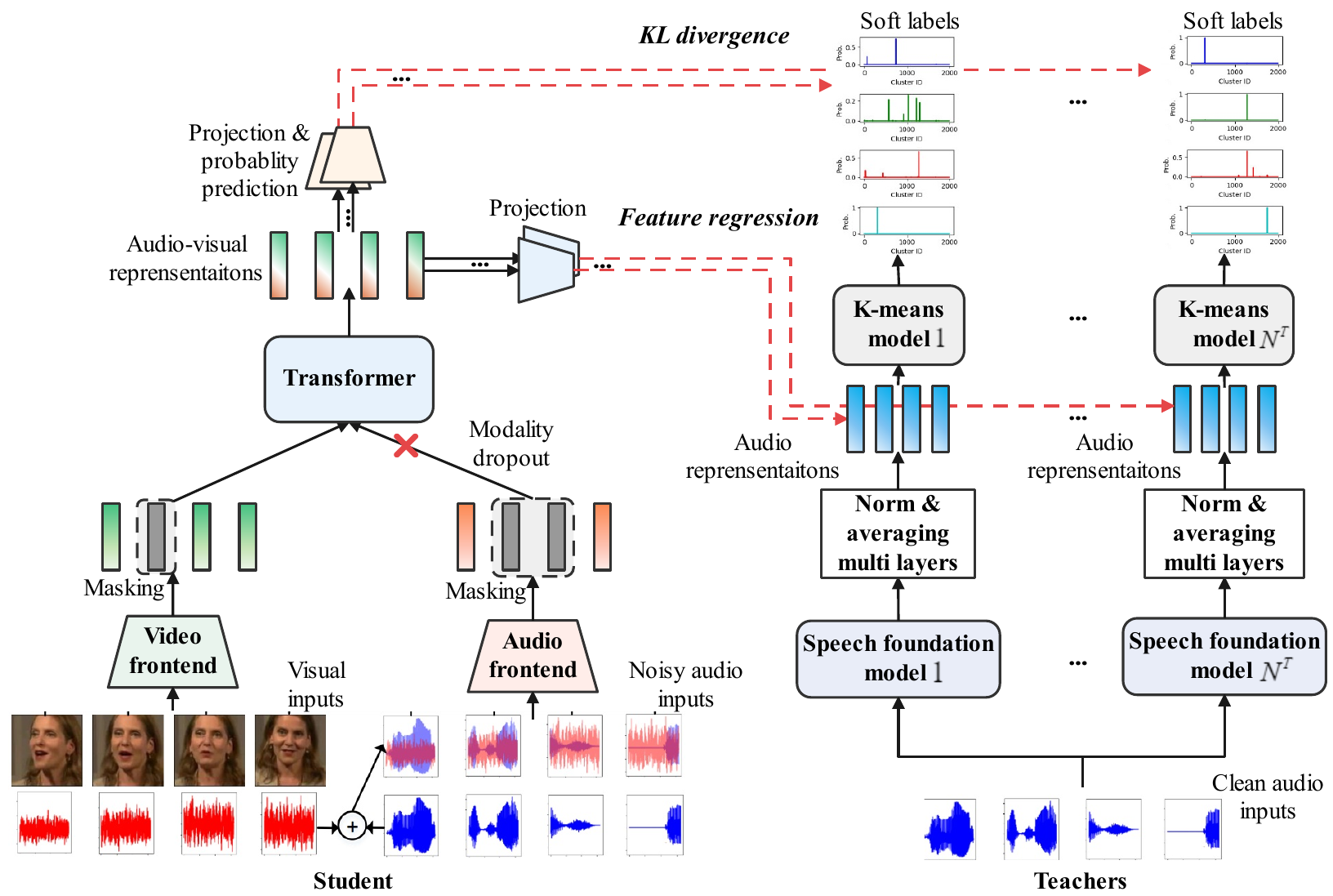}
    \caption{Illustration of our proposed method during the pretraining phase.}
    \label{fig:figure2}
\end{figure}

In our method, speech foundation models (SFMs) serve as teachers, whose hidden representations are extracted as learning targets for the student model. During pretraining phase, the audio-visual student model is optimized using our proposed  knowledge distillation loss function on unlabeled data.
During finetuning phase, KD loss is retained in conjunction with the primary training objective for downstream tasks. Figure~\ref{fig:figure1} illustrates the overall scheme of our proposed method.


\subsection{Student}
\label{sec:student}

The structure of the student is 
depicted in the left portion  of Figure~\ref{fig:figure2}.
This model processes both audio and visual inputs concurrently.
For the audio stream, Mel filterbank features are used. The visual stream incorporates the entire face images, departing from the conventional approach of using only the lip region-of-interest (ROI). Previous research~\cite{zhangbeyondlip}, as well as our own studies~\cite{Zhang2022face,Zhang2023av2vec} have demonstrated that entire faces contain more detailed information than lips alone, 
thereby enhancing visual representation learning.
To enhance noise robustness, the audio input of the student model, $\boldsymbol{A}$, is derived by adding random noise to the clean audio with a probability $p_{noise}$. 
Both audio and visual inputs are processed through modality-specific frontend to generate audio local features $\boldsymbol{F}^a \in \mathbb{R}^{T \times D_f}$ and 
 visual local features $\boldsymbol{F}^v \in \mathbb{R}^{T \times D_f}$ as
\begin{equation}
    \boldsymbol{F}^a = \text{AF}(\boldsymbol{A}), \boldsymbol{F}^v = \text{VF}(\boldsymbol{V}),
\end{equation}
where AF($\cdot$) and VF($\cdot$) represent audio and visual frontend respectively.
 Subsequently, features from each modality undergo independent corruption using a span-based mask. 
 Let $M_a$ denote the set of masked indices for audio.
 The corrupted audio feature 
 is derived as $\tilde{\boldsymbol{F}}^a
 = r(\boldsymbol{F}^a, M_a)$, where
 $\boldsymbol{F}^a_i$ is replaced with an embedding $\boldsymbol{e}^a$
if $i \in M_a$, with $i$ represents frame index.
 An analogous strategy is applied to corrupt visual features, utilizing different mask indices $M_v$ and embedding $\boldsymbol{e}^v$.
 Then modality dropout~\cite{shiavhubert} 
 is implemented to train the student 
 in the absence of either audio or video stream randomly. Specifically, with a probability
 $p_m$, both modalities are used. The probability of selecting audio is
 $p_a$ when only one modality is used, and features of the omitted 
 modality are set to zero.  The audio-visual features
 are then concatenated along the channel dimension as $\boldsymbol{F}^{av} = \text{concat}(\tilde{\boldsymbol{F}}^a, \tilde{\boldsymbol{F}}^v) \in \mathbb{R}^{T \times 2D_f}$.
 $\boldsymbol{F}^{av}$ is  subsequently  processed through a stack of Transformer blocks to generate contextualized audio-visual representations as
\begin{equation}
   \boldsymbol{O} = \text{Transformer}(\boldsymbol{F}^{av}),
\end{equation}
where
$\boldsymbol{O}=[\boldsymbol{o}_1, \dots, \boldsymbol{o}_T] \in \mathbb{R}^{T \times D_{o}}$ and  $D_{o}$  represents the channel dimension.

\subsection{Teacher}

Speech foundation models (SFMs) serve as the teachers in our proposed method.
For distilling student, latent representations from the last $k$ layers of the SFM are extracted from clean audio as 
\begin{equation}
    \{\boldsymbol{H}^{L-k}, \dots,  \boldsymbol{H}^L\}= \text{SFM}(\boldsymbol{A}^c),
\end{equation}
where $L$ denotes the total number of layers in the SFM, $\boldsymbol{H}^{i} \in \mathbb{R}^{T \times D_{h}}$, and $D_{h}$ represents the channel dimension of the SFM. Multi-layer representations are
first instance normalized and then averaged to obtain the final
teacher representations $\boldsymbol{H}^\mathcal{T} = [\boldsymbol{h}^\mathcal{T}_1, \dots, \boldsymbol{h}^\mathcal{T}_T] \in \mathbb{R}^{T \times D_h}$ as
\begin{equation}
    \boldsymbol{H}^{\mathcal{T}} = \frac{1}{k} \sum_{i=L-k}^{L}{\text{IN}(\boldsymbol{H}^i)},
\label{eq:eq4}
\end{equation}
where $\text{IN}( \cdot )$ represents instance normalization.

Research has shown that self-supervised SFMs encode different aspects of knowledge across various layers~\cite{pasad2021layer}. Consequently, averaging representations from multiple layers can enrich the target knowledge for more effective student distillation. Instance normalization is applied prior to averaging to prevent any single layer from dominating the combined representations.
Our experimental results in Section~\ref{sec:multilayer} confirmed the efficacy of multi-layer averaging for the self-supervised teacher. Interestingly, for the supervised teacher, using representations solely from the last layer yielded optimal results. This discrepancy likely reflects differences in the hierarchical encoding patterns of information, which arise from the distinct training objectives of self-supervised and supervised SFMs.



\subsection{Pretraining}

\subsubsection{Distillation loss function}

During the pretraining phase, the student is optimized using unlabeled audio-visual speech data through cross-modal knowledge distillation from SFM's representations.
The total knowledge distillation loss is a combination of a feature regression loss and a KL divergence loss,  as illustrated in Figure~\ref{fig:figure2}.
For feature regression loss, a linear projection layer is employed 
to the student's outputs to predict the teacher's representations as
\begin{equation}
L_\text{reg} = \frac{1}{T} \sum_{t=1}^T || \boldsymbol{W}\boldsymbol{o}_t - \boldsymbol{h}^\mathcal{T}_t||^2_2,
\end{equation}
where $\boldsymbol{W}$ are trainable parameters.

In AV-HuBERT~\cite{shiavhubert},
latent representations are
clustered using k-means algorithm, and then  
quantized 
by selecting the nearest cluster centroid 
to obtain pseudo-labels.
However, this approach loses the
hidden knowledge about the distances from current representation frame to other non-nearest
cluster centroids.
To extract richer knowledge, we propose constructing soft labels that capture the relationships between the current frame and all cluster centroids. 
Let the cluster centroids be  $\boldsymbol{c}_1 \dots \boldsymbol{c}_N$, where $N$ represents the total number of clusters. The $i$-th value of soft label $\boldsymbol{l}^\text{soft}_t \in \mathbb{R}^N$  for the $t$-th frame is computed as
\begin{equation}
\boldsymbol{l}^\text{soft}_t(i) = \frac{\text{exp}(-||\boldsymbol{h}^\mathcal{T}_t - \boldsymbol{c}_i||^2_2 / (\tau'I))}{\sum_{n=1}^N \text{exp}(-||\boldsymbol{h}^\mathcal{T}_t - \boldsymbol{c}_n||^2_2 / (\tau'I ))},
\end{equation}
where $I$ represents the inertia computed by k-means clustering, i.e., the sum of squared distances of samples to their closest cluster centriod. The hyper parameter $\tau'$ controls the smoothness of soft labels. The student is then trained using a 
KL divergence loss as 
\begin{equation}
L_\text{kld}= \frac{1}{T} \sum_{t=1}^{T} \text{KLD}(\text{softmax}(\cos(\boldsymbol{U}\boldsymbol{o}_t,\boldsymbol{E})/\tau), \boldsymbol{l}^\text{soft}_t),
\end{equation}
where $\boldsymbol{U}$ and $\boldsymbol{E}$ are trainable parameters, $\text{KLD}(\cdot)$  represents the KL divergence loss,
and $\text{cos}(\cdot)$ represents the cosine function.

The total knowledge distillation loss function incorporates both $L_\text{reg}$ and $L_\text{kld}$,  framing the problem as a multitask learning scenario.  To balance these two losses,
we employ a multitask learning method handles with multiple training objective functions,  specifically
Aligned-MTL-UB~\cite{senushkin2023independent}. In this approach, the gradients of each loss with respect to the student's outputs $\boldsymbol{O}$ are first computed. These gradients are then used to determine the weighting parameters for gradient aggregation.
Aligned-MTL aligns the orthogonal components
of the linear system of gradients, thereby enhancing the
training stability in multitask learning.
Unlike previous studies~\cite{shiavhubert,Zhang2023av2vec,Lian2023avdata2vec} that only compute loss only over masked regions, 
our method considers 
both masked and unmasked regions. This method facilitates more effective knowledge distillation, as demonstrated in our experiments in Section~\ref{sec:mask}.

\subsubsection{Multi-teacher ensemble}
Our method further proposes employing an ensemble of
multiple teachers to enhance the student's generalization capability.
Specifically, representations from multiple $N^{\mathcal{T}}$ teachers are extracted.
The student uses separate projection heads $\{\boldsymbol{W}_1, \boldsymbol{U}_1, \boldsymbol{E}_1\}$, 
$\dots$, $\{\boldsymbol{W}_{N^{\mathcal{T}}}, \boldsymbol{U}_{N^{\mathcal{T}}}, \boldsymbol{E}_{N^{\mathcal{T}}}\}$ to compute the knowledge distillation loss for each teacher.
These losses are  then combined to pretrain the student using the Aligned-MTL algorithm. 

\subsection{Finetuning}

During finetuning, a randomly initialized Transformer-based decoder is employed, with the 
student model serving as the encoder.
The encoder is frozen for the first $n_\text{freeze}$ updates, after which the encoder-decoder model is optimized jointly using the labeled data for the remaining updates.
In addition to the primary training objective of the task, our proposed knowledge distillation loss
is still retained as an auxiliary training objective. 
The knowledge distillation loss is first scaled by a factor of $\lambda$ then 
summed with the primary loss to produce the total loss  for finetuning.

\section{Experiments and results}

\subsection{Implementation details}\footnote{Our code will be available at \url{https://github.com/jxzhanggg/DistillAV}.}
\textbf{Datasets.} For pretraining, LRS3 dataset\footnote{\url{https://mmai.io/datasets/lip_reading/}} was utilized, which comprises approximately 433 hours of English audio-visual speech data. 
Additionally, we employed
a combination dataset of LRS3 and English-only version of VoxCeleb2~\cite{chung2018voxceleb2}, curated by Shi et al.~\cite{shiavhubert}, totaling about 1759 hours of audio-visual speech data. 
Our ablation studies were conducted exclusively on the LRS3 dataset.
For finetuning, we adopted two subsets of LRS3: a 30-hour subset and the full 433-hour training set.
A validation set of 1000 utterances was reserved, and the results  were reported on LRS3 test set.
We used the facial images provided in the original datasets without cropping to lip region-of-interest (ROI).
The MUSAN dataset\footnote{\url{https://www.openslr.org/17/}} was employed for noise augmentation. Following the protocol outlined in~\cite{DBLP:journals/corr/abs-2201-01763}, 
we sampled and added noise to the audio input of the student model during pretraining. Noise augmentation was also applied during finetuning for the AVSR task.
To evaluate our proposed method in AVSR under noisy conditions, we constructed test sets with various noise types, including music, natural sounds, speech, and babble, at signal-to-noise ratios (SNRs) of -10, -5, 0, 5, and 10 dB, respectively.



\textbf{Model configurations.}
In our experiments, we employed two speech foundation models: a self-supervised model, WavLM, and a supervised model, iFLYTEK-speech.
WavLM~\cite{chen2022wavlm}, based on a Transformer backbone, utilizes a training objective of predicting pseudo-labels of simulated noisy speech in masked regions. We specifically used WavLM-Large\footnote{\url{https://huggingface.co/microsoft/wavlm-large}}, which comprises 24 Transformer blocks, 317M parameters and was trained on 94,000 hours of speech data.
iFLYTEK-speech, an in-house model developed by iFLYTEK company, is built on a Conformer backbone and trained on large-scale multilingual speech data. It is designed for various speech-to-text tasks, primarily speech recognition and speech translation. This model employs an encoder-decoder architecture, of which only the encoder is utilized for extracting speech representations. iFLYTEK-speech contains 16 Conformer blocks, approximately 300M parameters and was trained on over 200,000 hours of proprietary speech data.


The student model architecture followed previous studies~\cite{shiavhubert,Zhang2023av2vec}, employing a feed-forward network for the audio frontend and a 3DCNN-ResNet18 network for the visual frontend. 
Two configurations were implemented for the Transformer encoder: a \textbf{Base} version with 12 layers and a \textbf{Large} version with 24 layers.
Their embedding dimension/feed forward dimension/attention heads were 768/3072/12 and 1024/4096/16, respectively. 
The WavLM model generates representations at a frame rate of 50 fps, which is double that of the student model's outputs. To address this discrepancy, we projected each student output frame onto two consecutive teacher frames when computing the knowledge distillation (KD) loss. In contrast, the frame rate of the iFLYTEK-speech encoder representations matches that of the student model.
The value of $k$ in Equation~\ref{eq:eq4} was set to 8 for the WavLM-based teacher and $1$ for the iFLYTEK-speech-based teacher. These values were optimized using our validation set, as detailed in Section~\ref{sec:multilayer}.

During pretraining,
80\% of audio features
and 30\% of video features were masked for the student. $p_m$ and
$p_a$ were both set to 0.5.
The temperature parameters $\tau$ and $\tau'$ were both set to 0.1.
The inertia values of k-means model for WavLM and iFLYTEK-speech representations
were 352.6 
and 19.1 
respectively.
The number of k-means clusters, $N$, was set to 2000.
During finetuning, a decoder Transformer with 6 layers was employed. Its embedding dimension/feed forward dimension/attention heads were 768/3072/4.
We utilized subword units with a vocabulary size of 1000 as the targets.
 The value of $n_\text{freeze}$ was optimized using our validation set. 
 Our experiments revealed that freezing the encoder throughout the fine-tuning process yielded the best performance when using the 30-hour labeled dataset. Consequently, the auxiliary KD loss for regularizing the encoder was not employed in this scenario.
 When fine-tuning with the larger 433-hour labeled dataset, we incorporated the auxiliary KD loss with a $\lambda$ value of 0.1.

 \textbf{Training costs and stability.}
 Unlike AV-HuBERT~\cite{shiavhubert} and u-HuBERT~\cite{hsu2022u}, our proposed method eliminates the need for iterative processing between feature clustering and mask prediction steps. The computational cost of our entire pretraining process is comparable to a single iteration of the AV-HuBERT model, resulting in significant reduction in training resources.
8 Tesla A100 GPUs were utilized in our experiments. For the 433-hour dataset, the pretraining process (400k steps) took approximately 5.3 days and 6.2 days for base and large versions, respectively. When extending to the 1759-hour dataset (800k steps), the pretraining time increased to 8.2 days and 10.3 days for base and large models. The finetuning process is notably efficient, completing within 16 hours.

Another category of AVSSL methods, such as our previously proposed AV2vec~\cite{Zhang2023av2vec}, requires an online-updated teacher module. These approaches typically necessitate techniques like feature normalization and careful selection of EMA hyperparameters to prevent target collapse~\cite{Zhang2023av2vec,haliassos2022jointly}. In contrast, our method leverages representations from well-established speech foundation models for knowledge distillation, effectively eliminating the risk of feature collapse. Our experimental results demonstrated robust training stability with no convergence issues.


\subsection{Comparison with baselines}

\begin{table}[]
    \centering
    \scalebox{0.9}{
    \begin{tabular}{c c c c c c c c}
     \hline
     \multirow{2}{*}{Methods} & Unlabeled  & Label & Encoder & \multirow{2}{*}{Criterion}  & \multirow{2}{*}{VSR} & \multirow{2}{*}{ASR} & \multirow{2}{*}{AVSR} \\
     & Data & Data & Size & & & & \\
     \hline
     \multicolumn{8}{c}{\emph{Self-supervised model (Base)}} \\
     \hline
    AV-HuBERT~\cite{shiavhubert}  &  \multirow{6}{*}{433h} & \multirow{6}{*}{30h} & 103M & CE & 51.8 & 4.9 & $4.7^2$ \\
    RAVen~\cite{haliassos2022jointly}  & & & 97M & CTC+CE & 47.0 & 4.7 & -- \\
    VATLM~\cite{zhu2023vatlm}$^1$  &  &  & 103M & CE & 48.0 & -- & 3.6 \\
    AV-data2vec~\cite{Lian2023avdata2vec}  &  &  & 103M  & CE & 45.2  & 4.4 & 4.2 \\
    AV2vec-MLM~\cite{Zhang2023av2vec}   &   & & 103M & CE & 39.4 & 5.6 & 5.4 \\
    AVFusion-Whisper~\cite{10448047}            & & & 87M & CE+KD & -- & -- & 2.8 \\
     Lip2Vec~\cite{djilali2023lip2vec}                                   &  &  & 240M  & CTC+KD & 49.5 & -- & -- \\
    \textbf{\emph{Proposed}} &   & & 103M & CE & \textbf{37.2} & \textbf{3.0} & \textbf{2.8} \\
    \hline
    AV-HuBERT~\cite{shiavhubert}  &  \multirow{5}{*}{1759h} & \multirow{5}{*}{30h} & 103M & CE & 46.1 & 4.6 & 4.0$^2$ \\
    RAVen~\cite{haliassos2022jointly}  & & & 97M & CTC+CE & 40.2 & 3.8 & -- \\
    VATLM~\cite{zhu2023vatlm}$^1$  &  &  & 103M & CE & 42.6 & -- & 3.4 \\
    AV-data2vec~\cite{Lian2023avdata2vec}  &  &  & 103M  & CE & 37.8  & 3.7 & 3.3 \\
     Lip2Vec~\cite{djilali2023lip2vec}                                   &  &    & 240M & CTC+KD & 40.6 & -- & -- \\
    \textbf{\emph{Proposed}} &   & & 103M & CE & \textbf{36.9} & \textbf{2.5} & \textbf{2.2} \\
    \hline
    \multicolumn{8}{c}{\emph{Self-supervised model (Large)}} \\
     \hline
        AV-HuBERT~\cite{shiavhubert}  &  \multirow{3}{*}{433h} & \multirow{3}{*}{30h} & 325M & CE & 44.8 & 4.5 & $4.2^2$ \\
    AV-data2vec~\cite{Lian2023avdata2vec}  &  &  & 325M  & CE & 40.5  & 3.7 & 3.4 \\
      AVFusion-Whisper~\cite{10448047}            &   &  & 257M & CE+KD & -- & --& 2.3 \\
     Lip2Vec~\cite{djilali2023lip2vec}                                   &  &    & 727M & CTC+KD & 55.4 & -- & -- \\
    \textbf{\emph{Proposed}} &   & & 325M & CE & \textbf{33.2} & \textbf{2.4} & \textbf{1.8} \\
    \hline
    AV-HuBERT~\cite{shiavhubert}  &  \multirow{5}{*}{1759h} & \multirow{5}{*}{30h} & 325M & CE & 32.5 & 2.9 & $3.3^2$ \\
    RAVen~\cite{haliassos2022jointly}  & & & 671M & CTC+CE & 33.1 & 2.6 & -- \\
    VATLM~\cite{zhu2023vatlm}$^1$  &  &  &  325M & CE & 31.6 & -- & 2.7 \\
    AV-data2vec~\cite{Lian2023avdata2vec}  &  &  &  325M  & CE & 30.8  & 2.7 & 2.7 \\
    Lip2Vec~\cite{djilali2023lip2vec}                                   &  &    & 727M & CTC+KD & 31.2 & -- & -- \\
    \textbf{\emph{Proposed}} &   & &  325M & CE & \textbf{30.2} & \textbf{2.2} & \textbf{2.1} \\
    \hline
    \end{tabular} 
    }
    \caption{Word error rate (WER) (\%) results on LRS3 test set with 30-hour labeled data. $^1$VATLM uses additional 3846h audio, 452h audio-text and 600M text data. $^2$Reproduction results from~\cite{Lian2023avdata2vec}.}
    \label{tab:tab2}
\end{table}

\begin{table}[]
    \centering
    \scalebox{0.87}{
    \begin{tabular}{c c c c c c c c}
    \hline
             \multirow{2}{*}{Methods} & Unlabeled  & Label & Encoder & \multirow{2}{*}{Criterion}  & \multirow{2}{*}{VSR} & \multirow{2}{*}{ASR} & \multirow{2}{*}{AVSR} \\
     & Data & Data & Size & & & & \\
     \hline
     \multicolumn{8}{c}{\emph{Supervised model}} \\
     \hline
     Afouras et al.~\cite{afouras2018deep} & -- & 1519h & -- & CE & 58.9 & 8.3 & -- \\
    Makino et al.~\cite{makino2019recurrent} & -- & 31kh &  -- & Transducer & 33.6 & 4.8 & 4.5 \\
    Prajwal et al.~\cite{prajwal2022sub} & -- & 2726h & -- & CE & 30.7 & -- & -- \\
    Serdyuk et al.~\cite{SerdyukBS22} & -- & 90kh & -- & Transducer & 17.0 & -- & 1.6 \\
    Chang et al.~\cite{10446532} & -- & 100kh & -- & Transducer & 12.8 & -- & 0.9 \\ 
    \hline
     \multicolumn{8}{c}{\emph{Self-supervised model (Base)}} \\
     \hline
     AV-HuBERT~\cite{shiavhubert} &  \multirow{5}{*}{433h} & \multirow{5}{*}{433h}  & 103M & CE & 44.0 & 3.0 & $2.8^2$ \\
     RAVen~\cite{haliassos2022jointly} & & & 97M & CTC+CE & 39.1 & 2.2 & -- \\
     AV-data2vec~\cite{Lian2023avdata2vec} & & & 103M & CE & 39.0 & 2.0 & \textbf{1.8} \\
      AV2vec-MLM~\cite{Zhang2023av2vec} & & & 103M & CE & 34.4 & 2.7 & 2.5 \\
      Lip2Vec~\cite{djilali2023lip2vec}                           & & & 240M & CTC+KD & 42.0 & -- & -- \\ 
     \textbf{\emph{Proposed}} & & & 103M & CE+KD & \textbf{34.3} & \textbf{1.9} & \textbf{1.8} \\
     \hline
     AV-HuBERT~\cite{shiavhubert} &  \multirow{5}{*}{1759h} & \multirow{5}{*}{433h} & 103M & CE & 34.8 & 2.0 & $1.8^2$ \\
      RAVen~\cite{haliassos2022jointly} & & & 97M & CTC+CE & 33.1 & 1.9 & -- \\
     VATLM~\cite{zhu2023vatlm}$^1$ &  &  &  103M & CE & 34.2 & -- & 1.7 \\
      AV-data2vec~\cite{Lian2023avdata2vec} &  &  &  103M & CE & 32.7 & \textbf{1.7} & \textbf{1.4} \\
      Lip2Vec~\cite{djilali2023lip2vec}                           & & & 240M & CTC+KD & 34.1 & -- & -- \\ 
      \textbf{\emph{Proposed}} & & & 103M & CE+KD & \textbf{31.4} & 1.9 & 1.7 \\
      \hline
      \multicolumn{8}{c}{\emph{Self-supervised model (Large)}} \\
      \hline
      AV-HuBERT~\cite{shiavhubert} &  \multirow{3}{*}{433h} & \multirow{3}{*}{433h} & 325M & CE & 41.6 & 2.7 & $2.5^2$ \\
    AV-data2vec~\cite{Lian2023avdata2vec} & & & 325M & CE & 37.4 & 1.9 & 1.7 \\
          Lip2Vec~\cite{djilali2023lip2vec}                           & & & 727M & CTC+KD & 50.1 & -- & -- \\ 
    \textbf{\emph{Proposed}} & & & 325M & CE+KD & \textbf{31.5} & \textbf{1.8} & \textbf{1.6} \\
    \hline
  AV-HuBERT~\cite{shiavhubert} &  \multirow{6}{*}{1759h} & \multirow{6}{*}{433h} & 325M & CE & 28.6 & \textbf{1.3} & $1.4^2$ \\
     RAVen~\cite{haliassos2022jointly} & & & 671M & CTC+CE & 28.2 & 1.4 & -- \\
     VATLM~\cite{zhu2023vatlm}$^1$ & & &  325M & CE & 28.4 & -- & \textbf{1.2} \\
     u-HuBERT~\cite{hsu2022u}$^1$ & & & 325M & CE & 27.2 & 1.4 & \textbf{1.2} \\
     AV-data2vec~\cite{Lian2023avdata2vec} & & & 325M & CE & 28.5 & 1.4 & 1.3 \\
       Lip2Vec~\cite{djilali2023lip2vec}                           & & & 727M & CTC+KD & \textbf{26.0} & -- & -- \\ 
      \textbf{\emph{Proposed}} & & & 325M & CE+KD & 26.2 & 1.4 & 1.3 \\
      \hline
    \end{tabular}
    }
    \caption{WER (\%) results on LRS3 test set with 433-hour labeled data. $^1$VATLM uses additional 3846h audio, 452h audio-text and 600M text data, and u-HuBERT uses additional 452h audio data. $^2$Reproduction results from~\cite{Lian2023avdata2vec}.}
    \label{tab:tab3}
\end{table}

\subsubsection{Results on clean test set}

Table~\ref{tab:tab2}
and Table~\ref{tab:tab3} summarize the results on the clean LRS3 test set for models finetuned with 30 and 433 hours of labeled data, respectively. 
When using low-resource labeled data (30 hours), our proposed method outperformed baselines across VSR, ASR, and AVSR tasks, as shown in Table~\ref{tab:tab2}. 
Notably, the superior performance in the VSR task demonstrates the effectiveness of cross-modal knowledge transfer from the audio to the visual modality.
When using high-resource labeled data (433 hours),
our method outperformed all self-supervised baselines in VSR task except Lip2vec pretrained with 1759h data, which achieved comparable performance, as shown in Table~\ref{tab:tab3}.
While the supervised baseline by Chang et al.~\cite{10446532} obtained the lowest word error rate (WER),  it utilized a substantial 
dataset of approximately 100,000 hours of audio-visual speech data.
For ASR and VSR tasks, our method outperformed or matched the baselines. However, AVSR results on the clean LRS3 test set
 have approached saturation, diminishing the significance of inter-model comparisons.
Therefore, 
we proceeded to evaluate AVSR performance of our proposed method on more challenging noisy test sets.

\subsubsection{AVSR results on noisy test sets}

\begin{figure}
    \centering
    \includegraphics[width=\linewidth]{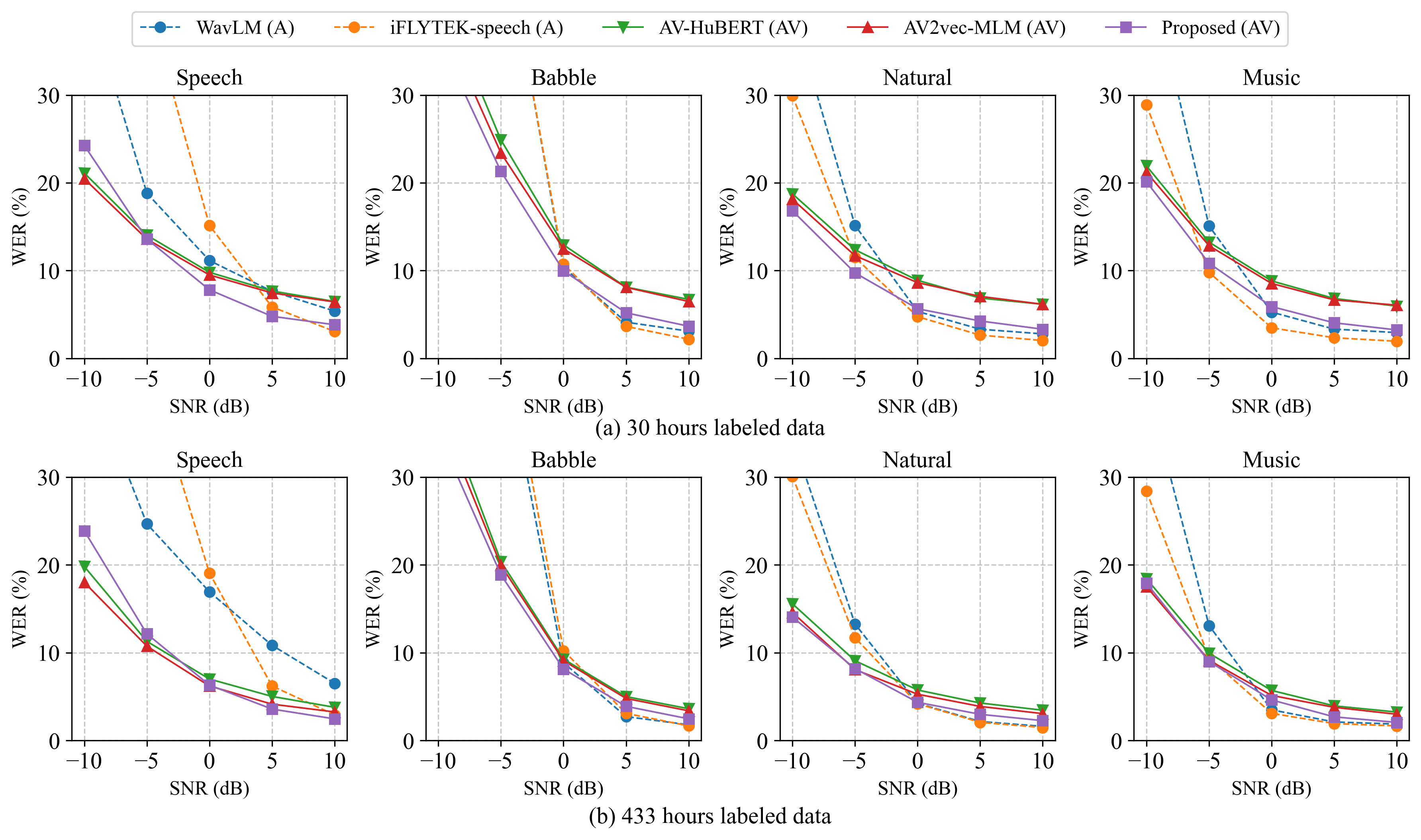}
    \caption{WER (\%) results on the test sets corrupted by various types of noise. ``A'' and ``AV'' represent models using audio and audio-visual inputs respectively.}
    \label{fig:fig2}
\end{figure}

We constructed four types of test sets by mixing various noise types (speech, babble, natural, and music) with the clean audio from the LRS3 test set. 
The teacher models, WavLM and iFLYTEK-speech,  were finetuned using 
audio-only inputs. Two AVSR baselines were adopted, including
AV-HuBERT~\cite{shiavhubert} and AV2vec-MLM~\cite{Zhang2023av2vec}. AV-HuBERT was reproduced using our configuration with face images as inputs~\cite{Zhang2023av2vec}. For pretraining, we utlized 433 hours of data from LRS3.
Figure~\ref{fig:fig2} summarizes the experimental results.
In low signal-to-noise ratio (SNR) scenarios, 
models with audio-visual inputs significantly  outperformed ASR models,
particularly on test sets with speech or babble noise.
However, as audio quality improved, the performance of speech foundation models increased dramatically, ultimately achieving the lowest WER under 10 dB SNR conditions across most noisy test sets.

When compared to AVSR baselines,
our proposed method demonstrated superior performance under most conditions, except for low SNR test sets with speech noise. 
This finding suggests that our method may face challenges of over-reliance on the audio modality when processing data corrupted by heavy speech noise. Orthogonal to this work, such issue in AVSR was addressed by MSRL~\cite{chen2023leveraging}, which could be utilized to further improve AVSR performance in future studies.

\subsection{Ablation studies}


\subsubsection{Multi-teacher ensemble}

\begin{table}[]
\centering
\scalebox{0.85}{
    \begin{tabular}{ c c c c c  c  c}
    \hline
        \multirow{2}{*}{Labeled Data} & \multicolumn{2}{c}{Teacher} & \multicolumn{2}{c}{Base} & \multicolumn{2}{c}{Large} \\
        \cline{2-3}\cline{4-7}
                          & WavLM & iFLYTEK-speech   & ASR &  VSR  & ASR & VSR \\
        \hline
        \multirow{3}{*}{30h} &   \CheckmarkBold & \XSolidBrush &  3.8   &   38.9  & 2.9   & 33.3 \\
                             &  \XSolidBrush & \CheckmarkBold  &  3.1   &  38.0    & 2.5   & 33.7 \\
                             &   \CheckmarkBold & \CheckmarkBold  &  \textbf{3.0}    &   \textbf{37.2} &   \textbf{2.4}  & \textbf{33.2} \\
        \hline
        \multirow{3}{*}{433h}  &   \CheckmarkBold & \XSolidBrush &  2.3   & 35.4    & 2.0   & \textbf{31.2} \\
                              &  \XSolidBrush & \CheckmarkBold  &    \textbf{1.9} &   34.7  &   1.9 & 32.9 \\
                             &   \CheckmarkBold & \CheckmarkBold & 
                            \textbf{1.9}    & \textbf{34.3}   &  \textbf{1.8}   & 31.5 \\
        \hline
    \end{tabular}
    }
    \caption{WER (\%) results on LRS3 test set of our proposed method using  WavLM, iFLYTEK-speech-based, and an ensemble of the both as the  teacher models.}
    \label{tab:tab1}
\end{table}

This section presents an ablation study of our proposed multi-teacher ensemble method. 
We used the same hyperparameters as those in previous section. 433-hour LRS3 dataset was used for pretraining.
We compared  the performance of using WavLM, iFLYTEK-Speech, and an ensemble of both as the teacher models. 
The experimental results are summarized in Table~\ref{tab:tab1}.
For the ASR task, the student distilled from iFYTEK-speech-based teacher achieved
a lower WER than the one distilled from WavLM. In the VSR task, the Base student model performed better when using iFLYTEK-speech as the teacher compared to WavLM. 
However, this trend was reversed for the Large student model. Notably, the ensemble of both teachers enhanced the generalization of knowledge distillation, resulting in more robust overall performance across both tasks.

\subsubsection{Knowledge distillation loss}
This section examines our proposed loss function for knowledge distillation. Experiments were conducted using only the iFLYTEK-speech teacher model and the Base version of the student. 
433-hour LRS3 dataset was used for pretraining.
 We compared our proposed  method with several variants 
for representational knowledge distillation. \textbf{Reg} represents
utilizing only the feature regression loss. \textbf{CE} represents employing cross-entropy loss with discrete pseudo-labels, as in AV-HuBERT~\cite{shiavhubert}. \textbf{KLD} represents using only the proposed KL divergence loss function with soft labels.  \textbf{Reg+CE} represents combining
cross-entropy and feature regression loss.
Our proposed method is denoted by \textbf{Reg+KLD}. The results are presented
in Table~\ref{tab:tab4}. Our proposed KLD loss using soft labels outperformed the CE loss using discrete labels, demonstrating its effectiveness in transferring the teacher's hidden knowledge.
When combined with feature regression loss, the student model achieved the lowest WER in both ASR and VSR tasks. 
During finetuning with 30 hours of labeled data, the encoder remained frozen, and the auxiliary KD loss was not applied. 
When finetuning with high-resource labeled data (433 hours), 
WER was further reduced for both ASR and VSR tasks, particularly ASR. 
These results 
demonstrated the effectiveness of our proposed KD loss during the both pretraining and finetuning phases.


\begin{table}[]
    \centering
    \scalebox{0.9}{
    \begin{tabular}{l l c c c c}
        \hline
        \multicolumn{2}{c}{Knowledge Distillation Loss} & \multicolumn{2}{c}{30h} & \multicolumn{2}{c}{433h} \\\
        Pretrain &  Finetune &  ASR & VSR & ASR & VSR \\
        \hline
        Reg & -- & 3.2 & 38.5 & 2.4 & 35.6 \\
        CE & -- & 3.4 & 40.2 & 2.5 & 36.6 \\
        KLD & -- & 3.3 & 38.1 & 2.4 & 34.9 \\
        Reg+CE & -- & 3.2 & 38.9 & 2.5 & 35.4 \\
        Reg+KLD (\emph{proposed}) & -- & \textbf{3.1} & \textbf{38.0} & 2.4 & 34.8 \\
        \hline
        \multirow{5}{*}{Reg+KLD (\emph{proposed})} & Reg & -- & --  & 2.0 & 34.9 \\
        & CE & -- & -- & 2.0 & 34.9 \\
        & KLD & -- & -- & 2.0 & 34.8 \\
        & Reg + CE & -- & -- & 2.1 & 35.1 \\
        & Reg + KLD (\emph{proposed}) & -- & -- & \textbf{1.9} & \textbf{34.7} \\
        \hline
    \end{tabular}
    }
    \caption{WER (\%) results from the ablation study of our proposed
    knowledge distillation loss. ``Reg'', ``CE'', and ``KLD'' represent feature regression, cross-entropy loss using discrete pseudo labels, and KL divergence loss using our proposed soft labels, respectively.}
    \label{tab:tab4}
\end{table}

\subsubsection{Model inputs}

\begin{table}[]
    \centering
    \scalebox{0.9}{
    \begin{tabular}{c c c c c c c}
    \hline
    \multicolumn{2}{c}{Inputs}  & \multirow{2}{*}{Video Type} & \multicolumn{2}{c}{30h} & \multicolumn{2}{c}{433h} \\
    Audio & Video & & ASR & VSR & ASR & VSR \\
    \hline
    \multicolumn{7}{c}{Teacher (iFLYTEK-ASR)} \\
    \hline
        \CheckmarkBold & \XSolidBrush & -- &  1.7 & -- & 1.4 & -- \\
    \hline
    \multicolumn{7}{c}{Student} \\
    \hline
    \CheckmarkBold & \XSolidBrush & -- &  \textbf{3.0} & -- & \textbf{2.4} & -- \\
     \XSolidBrush & \CheckmarkBold & Lip &  -- & 44.4 & -- & 41.7 \\
    \CheckmarkBold &  \CheckmarkBold  & Lip & \textbf{3.0} & 40.8 & 2.5 & 38.6 \\
    \CheckmarkBold &  \CheckmarkBold & Face  &  3.1 & \textbf{38.0} & \textbf{2.4} & \textbf{34.8} \\
    \hline
    \end{tabular}
    }
    \caption{WER (\%) results from the ablation study of model inputs. ``Lip'' and ``Face'' indicate the use of lip region-of-interest and entire face images as visual inputs, respectively. 
    To ensure comparability, we equipped the teacher model with a randomly initialized decoder that shared the same configuration as the student's.}
    \label{tab:tab5}
\end{table}

This section compares models accepting different types of inputs, including 
an audio-only model, a visual-only model, and our proposed joint audio-visual model. Experiments were conducted using only the iFLYTEK-speech teacher model and the Base version of the student, with the knowledge distillation loss omitted during finetuning.
The experimental results are presented in Table~\ref{tab:tab5}.
The joint audio-visual model achieved  equal or better performance compared to the single-modality baselines. 
Particularly, 
it significantly improved VSR results, demonstrating the effectiveness of jointly modeling audio and video for cross-modal distillation.
We also compared the use of lip ROI versus full face as inputs. The face-based model significantly outperformed the lip ROI-based model in the VSR task, consistent with our previous research~\cite{Zhang2022face}.


\begin{table}[]
    \centering
    \scalebox{0.9}{
    \begin{tabular}{c c c c c c}
    \hline
    \multirow{2}{*}{Masking} & \multirow{2}{*}{KD Region} & \multicolumn{2}{c}{30h} & \multicolumn{2}{c}{433h} \\
     &   & ASR & VSR & ASR & VSR \\
    \hline
    \XSolidBrush  & All &  3.5 & 43.0  & 2.7 & 37.2 \\
    \CheckmarkBold  & Mask & 3.2 & 38.3  & 2.5 & 35.3 \\
    \CheckmarkBold & All & \textbf{3.1} & \textbf{38.0} & \textbf{2.4} & \textbf{34.8} \\
    \hline
    \end{tabular}
    }
    \caption{WER (\%) results from the ablation study of masking strategy. ``Masking'' indicates whether the span-based feature mask was applied to the student during pretraining. ``All'' and ``Mask'' represent
    computing KD loss over entire region and only the masked region, respectively.}
    \label{tab:tab6}
\end{table}

\subsubsection{Masking strategy}
\label{sec:mask}
This section investigates the
masking strategy used for the student during pretraining.
Experiments were conducted using only the iFLYTEK-speech teacher model and the Base version of the student, with the knowledge distillation loss omitted during finetuning.
 433-hour LRS3 dataset was used for pretraining.
The results are summarized in Table~\ref{tab:tab6}.
Applying span masking to the intermediate features
enhanced the model performance by 
compelling the model to utilize contextual information for recovering representations in masked regions.
This approach has been extensively explored in previous research~\cite{shiavhubert,Zhang2023av2vec,haliassos2022jointly}.
Unlike AV-HuBERT, which predicts only masked regions, our experiments demonstrated that
computing KD loss over all regions yielded the best performance. 
This may be attributed to the high-quality learning targets derived from SFMs, which provided accurate and useful knowledge across all regions.
\begin{figure}[t]
    \centering
    \subfigure[WavLM]{
    \begin{minipage}[b]{0.46\textwidth}
    \includegraphics[width=\textwidth]{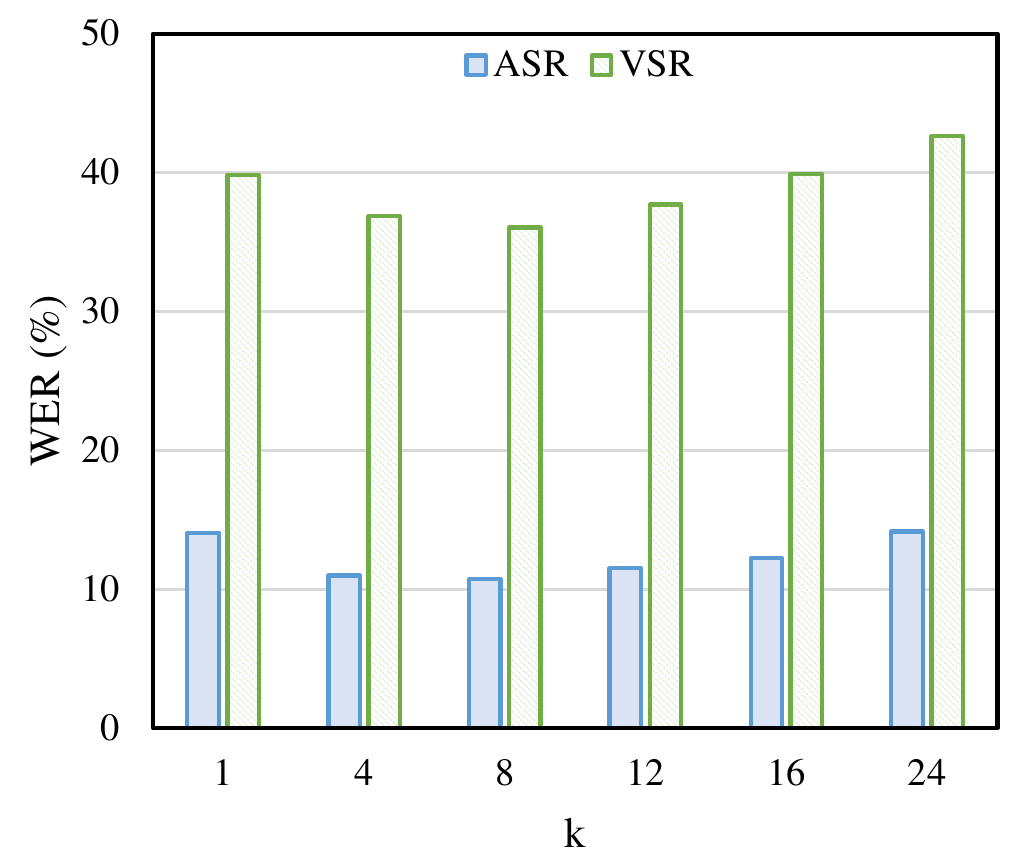}
    \end{minipage}}
     \subfigure[iFLYTEK-speech]{
    \begin{minipage}[b]{0.46\textwidth}
    \includegraphics[width=\textwidth]{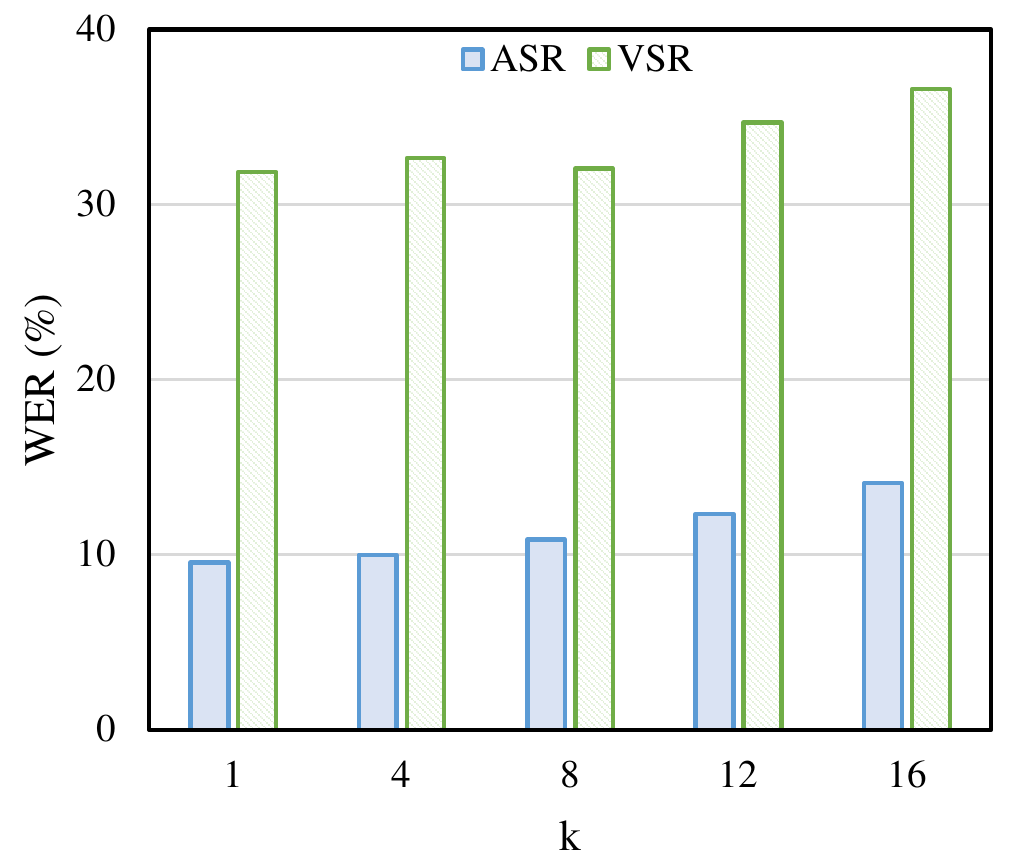}
    \end{minipage}}
    \caption{WER (\%) results on our validation set 
    as a function of $k$ (number of last teacher layers averaged) for WavLM and iFLYTEK-speech based teachers.
    }
    \label{fig:figure3}
\end{figure}

\subsubsection{Multi-layer representations aggregation}
\label{sec:multilayer}

\begin{figure}
    \centering
    \includegraphics[width=\linewidth]{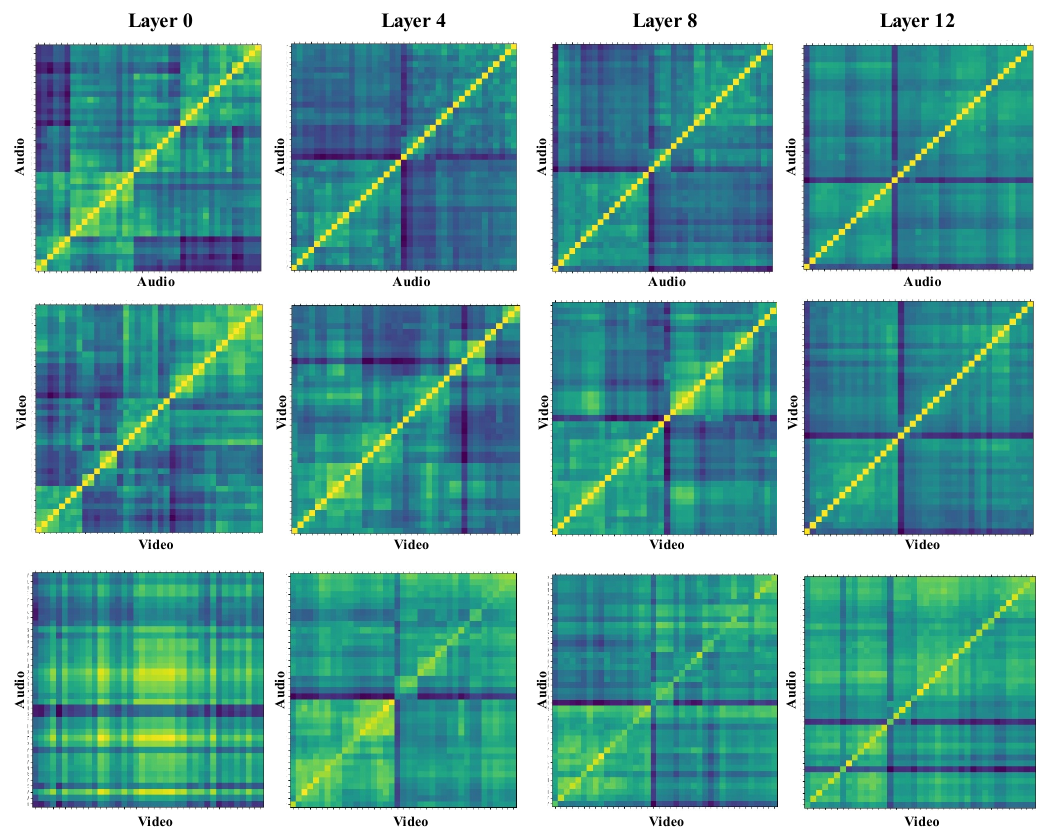}
    \caption{Distance matrices between phonemes using their corresponding representations from
    the student. Hierarchical clustering was applied to cluster phonemes based on their distances.
    ``Audio'' and ``Video'' refer to phoneme representations extracted from audio and visual inputs respectively. ``Layer N'' indicates representations from the N-th layer of the student model.}
    \label{fig:figure4}
\end{figure}

This section explores the effect of layer averaging of multi-layer representations
from the teacher model. Both WavLM and iFLYTEK-speech-based teacher model were investigated and compared, with the Base version of
the student model. 
Each model was pretrained for 200k updates.  433-hour and 30-hour LRS3 dataset were used for pretraining and finetuning, respectively. Knowledge distillation loss was not used during finetuning.
The results on our validation set are presented in Figure~\ref{fig:figure3}.
For the WavLM-based teacher,  the WER initially decreased but then increased as the number of averaged layers grew.
 Eventually, averaging the last 8 layers achieved  the lowest WER, 
 demonstrating the effectiveness of aggregating multi-layer representations for distillation.
In contrast, for the iFLYTEK-speech-based teacher model, the WER consistently increased as the number of averaged layers increased.
This divergent behavior between the two teachers can be attributed to their distinct methods of encoding knowledge across layers, which are influenced by their respective training objectives~\cite{pasad2021layer}.

 
\subsection{Representations visualization}

\begin{figure}
     \centering
     \includegraphics[width=0.9\linewidth]{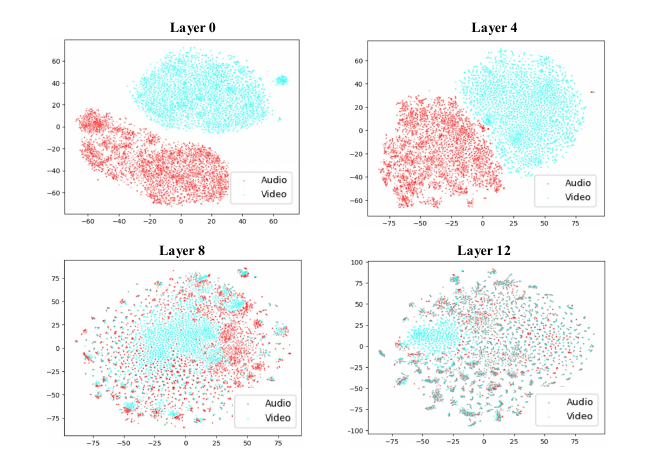}
     \caption{t-SNE visualization of audio and visual representations from the student model. Layer N” indicates representations from the N-th layer of the student model.
}
     \label{fig:figure5}
\end{figure}

In this section, our proposed model with ``Base'' configuration and
an ensemble of two teachers was used to extract the hidden representations from either audio or visual inputs.
The Montreal Forced Aligner tool\footnote{\url{https://montreal-forced-aligner.readthedocs.io/en/latest/index.html}} was then adopted to
align the speech with the phoneme sequences using test set, where
phoneme sequences was obtained via a grapheme-to-phoneme (G2P) model.
A phoneme set of 39 was utilized.
Hidden representations for each phoneme was collected and then averaged.
Using these representations, Euclidean distances between each pairs of phonemes were computed. A hierarchical clustering algorithm was subsequently 
employed to cluster the phonemes according to their distances.
The distance matrices are visualized in Figure~\ref{fig:figure4}.
As expected, audio representations exhibit better discriminability for phonemes compared to visual ones.
Interestingly, the visual representations in the final layer showed clear separation among different phonemes, which explains our model's strong performance in the VSR task.
The last row of this figure shows
that the audio and visual representations of the same phoneme became closer
 in deeper layers, indicating that both modality shared the same 
 space in deeper layers.
 Audio and visual representations were also projected into a two-dimensional space using the t-SNE algorithm~\cite{van2008visualizing} and visualized in Figure~\ref{fig:figure5}. 
This figure further confirms that the two distinct modality 
 inputs were progressively  processed and encoded within a unified space.
 These results align with recent studies demonstrating a trend towards representational convergence across modalities~\cite{wang2023image}.


\section{Conclusion}
In this work, 
we propose
an audio-visual representation learning method through cross-modal
distillation using speech foundation models. 
Our method requires only unlabeled data during pretraining.
Multi-layer representations from multiple speech foundation models are extracted to distill the student model. Our proposed knowledge distillation loss
combines a feature regression loss and a KL divergence loss with designed 
soft labels. The distillation loss is also applied during finetuning 
to enhance the performance of downstream tasks.
By distilling from speech foundation models, our method  effectively transfers the rich knowledge acquired from a vast volume of audio data to the video modality. 
We used two teacher models in our experiments: a self-supervised model, WavLM, and a supervised model, iFLYTEK-speech. 
The experimental results demonstrated that our proposed method 
 achieved superior or at least comparable performance to  baselines from previous studies.
Ablation studies confirmed the effectiveness of several key components of our 
method, including the multi-teacher ensemble, the proposed distillation loss, and the joint modeling of audio and visual modalities.
Furthermore, visualization of learned representations indicated that our model effectively encodes audio and visual modalities within a unified space. 

Our work has several limitations that remain to be addressed 
in future research.
First, in the lipreading task, while our method achieves lower WER than self-supervised learning baselines, it still falls short of the current best supervised models that utilize a large amount of labeled audio-visual data. This underscores the critical role of dataset scale in achieving optimal lipreading performance. As a future direction, we plan to collect more extensive audio-visual datasets to train lipreading models.
Second, in the AVSR task, although our model generally outperforms baselines, its performance deteriorates when the input audio is corrupted with speech noise at -10dB SNR. 
This observation may suggest that our model has an over-reliance on the audio modality.
Given that we adopted a conventional channel-wise concatenation strategy for audio and visual fusion, which was not the primary focus of this work, we intend to further investigate this issue.
In future work, we aim to design more flexible and effective methods for integrating audio and visual modalities to achieve more robust audio-visual speech recognition.



\section{Acknowledgements}
This work was supported by the National Natural Science Foundation of China (Grant No. 62401348), Fundamental Research Funds for the Central Universities (Grant No. GK202406005).

\section{Declaration of Generative AI and AI-assisted technologies in the writing process}

During the preparation of this work, the authors used iFLYTEK's Xinghuo\footnote{\url{https://xinghuo.xfyun.cn/}} in order to improve readability and language. After using this tool, the authors reviewed and edited the content as needed and take full responsibility for the content of the publication.

\bibliographystyle{elsarticle-num} 
\bibliography{ref}

\end{document}